\pdfoutput=1

\documentclass[pra,aps,superscriptaddress,showpacs,reprint]{revtex4-1}
\usepackage{graphicx}
\usepackage{subfigure}
\usepackage{amsmath}
\usepackage{amssymb}
\usepackage{dsfont}

\usepackage{color} % use colored text in latex

\usepackage{hyperref}
\hypersetup{
colorlinks=true,final=true,
        linkcolor=blue,
        citecolor=blue,
        filecolor=blue,
        urlcolor=blue,
        pdfauthor={L. Steffen},
        pdftitle={Realization of Deterministic Quantum Teleportation with Solid State Qubits}
}

\newcommand{\ket}[1]{| #1 \rangle}

\begin{document}

\title{Realization of Deterministic Quantum Teleportation with Solid State Qubits}

\author{L.~Steffen}
\altaffiliation{These authors contributed equally to this work.}
\author{A.~Fedorov}
\altaffiliation{These authors contributed equally to this work.}
\author{M.~Oppliger}
\author{Y.~Salathe}
\author{P.~Kurpiers}
\author{M.~Baur}
\author{G.~Puebla-Hellmann}
\author{C.~Eichler}
\author{A.~Wallraff}
\affiliation{Department of Physics, ETH Zurich, CH-8093 Zurich, Switzerland}

\date{\today}
\begin{abstract}
Transferring the state of an information carrier from a sender to a receiver is an essential primitive in both classical and quantum communication and information processing. In a quantum process known as teleportation the unknown state of a quantum bit can be relayed to a distant party using shared entanglement and classical information. Here we present experiments in a solid-state system based on superconducting quantum circuits demonstrating the teleportation of the state of a qubit at the macroscopic scale. In our experiments teleportation is realized deterministically with high efficiency and achieves a high rate of transferred qubit states. This constitutes a significant step towards the realization of repeaters for quantum communication at microwave frequencies and broadens the tool set for quantum information processing with superconducting circuits.
\end{abstract}

\pacs{}

\maketitle

\section{Introduction}
%Intro: How teleportation works
Engineered man-made macroscopic quantum systems based on superconducting electronic circuits are extremely attractive for experimentally exploring diverse questions in quantum information science \cite{Clarke2008,Houck2012}. At the current state of the art quantum bits (qubits) are fabricated, initialized, controlled, read-out and coupled to each other in simple circuits to realize basic logic gates \cite{Steffen2006a,Fedorov2012}, to create complex entangled states \cite{DiCarlo2010,Neeley2010a}, to demonstrate algorithms \cite{DiCarlo2009,Lucero2012} and error correction \cite{Reed2012}. The continuous improvement of coherence properties \cite{Paik2011} points at a promising future for developing quantum technology based on superconducting circuits.

A central challenge that will become increasingly important in future setups is the integration of many components in networks with arbitrary connecting topology on a planar chip. For the experiments discussed in this work we have developped a chip-based superconducting circuit architecture \cite{Wallraff2004,Mariantoni2011a} which uses cross-overs allowing to create such networks. In this architecture superconducting qubits are coupled to one or two resonators at the intersections of a set of vertical and horizontal transmission lines forming a grid~\cite{Helmer2009a}. This feature provides qubit/qubit coupling  across horizontal and vertical resonators. In addition, high fidelity single-shot read-out of single and joint two-qubit states is realized in dispersive measurements making use of low-noise parametric amplifiers~\cite{Castellanos2008,Yurke2006}. We demonstrate the power of such an architecture by realizing a deterministic quantum teleportation protocol, where all these ingredients of the circuit quantum electrodynamics toolbox are exploited. At a rate of $40\cdot10^3~\mathrm{events/s}$, exceeding many reported implementations of teleportation, quantum states are teleported over a distance of $6~\mathrm{mm}$ between two macroscopic quantum systems. The teleportation process is demonstrated to succeed with order unit probability for any input state due to our ability to deterministically prepare maximally entangled two-qubit states on demand as a resource and distinguish all four two-qubit Bell states in a single measurement during teleportation.

\section{Teleportation}

Quantum teleportation describes the process of transferring an unknown quantum state between two parties at two different physical locations making use of the non-local correlations provided by an entangled pair shared between the two and the exchange of classical information \cite{Bennett1993}.
This concept plays a central role in extending the range of quantum communication using quantum repeaters \cite{Gisin2002, Briegel1998} and can also be used to implement logic gates for universal quantum computation \cite{Gottesman1999}.

\begin{figure}[!t]
\includegraphics[width=0.48\textwidth]{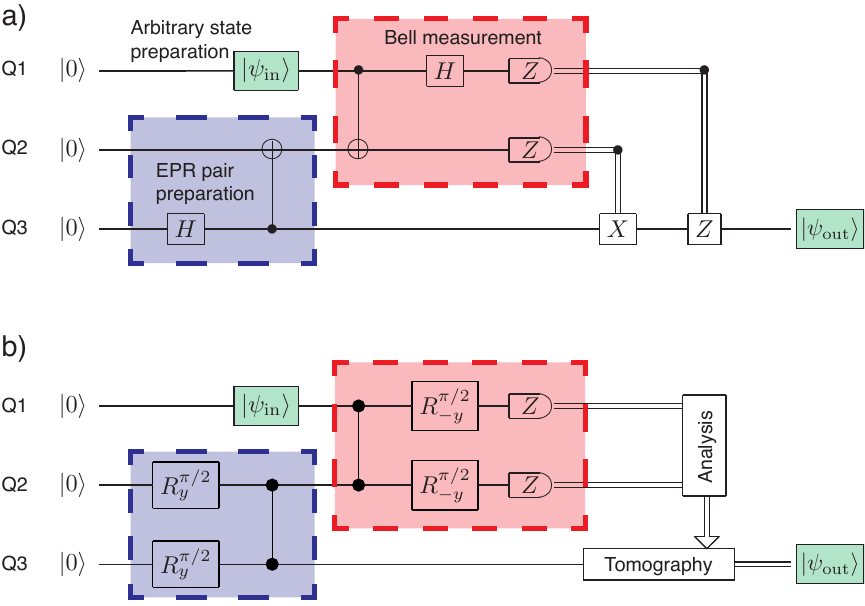}
\caption{a) Standard protocol of quantum teleportation. The protocol starts with the preparation of a Bell state between Q2 and Q3 (blue box) followed by the preparation of an arbitrary state $\ket{\psi_\mathrm{in}}$ (green box) and a Bell state measurement of Q1 and Q2 (red box). The classical information extracted by the measurement of Q1 and Q2 is transferred to the receiver to perform local gates conditioned on the measurement outcomes. After the protocol Q3 is in a state $\ket{\psi_\mathrm{out}}$ which ideally is identical to $\ket{\psi_\mathrm{in}}$ (also colored in green). Here, $H$ is the Hadamard gate, $X$ and $Z$ are Pauli matrices $\hat\sigma_x$ and $\hat\sigma_z$, respectively. The \textsc{cnot}-gate is represented by a vertical line between the control qubit ($\bullet$) and the target qubit ($\oplus$).
b) The protocol implemented in the experiment presented here uses controlled-\textsc{phase} gates indicated by vertical lines between the relevant qubits ($\bullet$),
and single qubit rotations $R_{\pm y}^\theta$ of angle $\theta$ about the $\pm y$-axis.}
\label{fig:pulses}
\end{figure}

In the original teleportation protocol \cite{Bennett1993}, the unknown state $\left|\psi_\mathrm{in}\right\rangle$ of qubit Q1 in possession of the sender is transferred to the receiver's qubit Q3, see \ Fig.~\ref{fig:pulses} a). To enable this task, sender and receiver prepare in advance a maximally entangled (Bell) state between an ancillary qubit Q2 and Q3. Then the sender performs a measurement of Q1 and Q2 in the Bell basis which projects the two qubits in his possession onto one of the four possible Bell states $\ket{\Phi^\pm} = \left(\ket{00}\pm\ket{11}\right)/\sqrt{2}$ and $\ket{\Psi^\pm} =\left(\ket{01}\pm\ket{10}\right)/\sqrt{2}$.
As a consequence the receiver's qubit Q3 is projected instantaneously onto a state $\left|\psi_\mathrm{out}\right\rangle = \left\{\mathds{1},\hat{\sigma}_x,\hat{\sigma}_z,i\hat{\sigma}_y\right\}\left|\psi_\mathrm{in}\right\rangle$,
which differs from the input state $\left|\psi_\mathrm{in}\right\rangle$ only by a single qubit rotation, depending on the four possible measurement results.
To always recover the original state $\left|\psi_\mathrm{in}\right\rangle$ the receiver can rotate the output state of Q3 conditioned on the outcome of the Bell measurement communicated to the receiver as two bits of information via a classical channel. This final step is frequently referred to as feed-forward, since the outcome of a measurement performed on one part is used to control the other part of the same quantum system. This is in contrast to acting back on the same quantum system in a feed-back process.

The success of the teleportation protocol in every instance with unit fidelity is counterintuitive from a classical point of view. The receiver's qubit does not interact with any other qubit after $\left|\psi_\mathrm{in}\right\rangle$ is prepared. The classical information sent by the sender is not sufficient to perfectly recreate $\left|\psi_\mathrm{in}\right\rangle$ at the receiver. Indeed assuming no entanglement between sender and receiver one can replicate the sender's state at best with a fidelity of $2/3$~\cite{Massar1995} since only a fraction of information about $\left|\psi_\mathrm{in}\right\rangle$ is obtained by a single projective measurement. Even if it were in principle possible to acquire complete knowledge of the state in a single measurement, the reconstruction of the state by sharing 2 bits of classical information would succeed only with $87\,\%$  fidelity~\cite{Gisin1996}.

%Intro: What has been done so far
In pioneering work the teleportation protocol was first implemented with single photons~\cite{Bouwmeester1997} over lab-scale distances and later also over km-scale distances in free space~\cite{Marcikic2003,Yin2012,Ma2012}. However, in these experiments only two out of four Bell states are distinguished unambiguously limiting the efficiency of the protocol to 50\,\% at best.
A proof of principle experiment which can distinguish all four Bell states was implemented using non-linear photon interaction~\cite{Kim2001} but the efficiency of the detection step is much below $1\,\%$.
With photonic continuous-variable states teleportation has been achieved deterministically for all measurement outcomes and the final conditional rotation has been implemented to complete the teleportation protocol~\cite{Furusawa1998, Lee2011}.
In atomic qubits fully deterministic quantum teleportation has been realized over micrometer scale distances with ions in the same trap~\cite{Riebe2004, Barrett2004}. Non-deterministically the protocol has also been implemented between ions in different traps \cite{Olmschenk2009} and in atomic ensembles~\cite{Bao2012a}.
Using nuclear magnetic resonance techniques for spin ensembles a teleportation-like protocol was implemented over interatomic distances by replacing the read-out and feed-forward step with dephasing and conditioned unitary operations~\cite{Nielsen1998}.

\begin{figure*}[!t]
\includegraphics[width=0.98\textwidth]{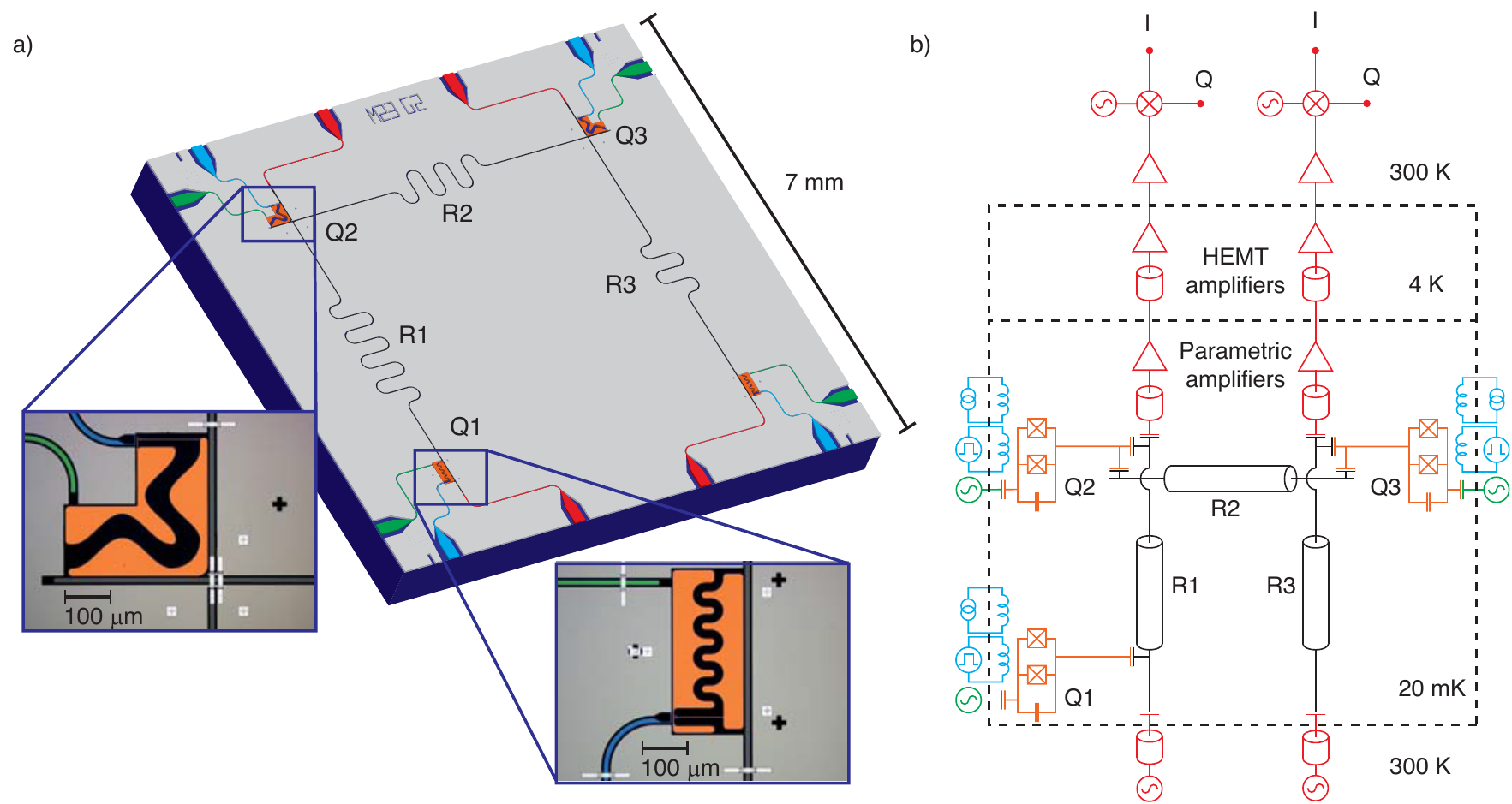}
\caption{a) Rendered image of the chip-design containing resonators R1, R2 and R3 (black) with corresponding in- and output lines (red) used for read-out and coupling of three transmon qubits Q1, Q2 and Q3 (orange). The fourth qubit in the lower right corner of the chip is not used in the presented experiments. The local microwave lines (green) are used for single qubit rotations while the local flux-bias lines (blue) allow for nanosecond time control of the qubit frequencies to implement two-qubit operations. The insets show a false-color micrograph of Q1 (bottom-right) coupled to resonator R1 and a false-color micrograph of Q2 (left) coupled to resonators R1 and R2. The aluminum airbridges visible as bright white strips realize cross-overs for the resonator lines which enhances scalability of this planar architecture. Airbridges are also used to suppress spurious electromagnetic modes by connecting the ground planes across the coplanar wave guides. b) Simplified schematic of the measurement setup with the same color code as in a), for details see text.}
\label{fig:schematicsample}
\end{figure*}

%Intro: What we do in the presented experiments
%Here we demonstrate an implementation of quantum teleportation using macroscopic superconducting circuits as our quantum systems. We use an architecture in which the superconducting qubits are positioned at the intersections of a set of vertical and horizontal resonators forming a grid of transmission lines~\cite{Helmer2009a}. This approach provides selective qubit measurement and pairwise qubit coupling.
%In addition using parametric amplifiers~\cite{Yurke2006}, we are able to distinguish all four Bell states in a single measurement and show by tomographical reconstruction that the receiver's qubit is projected onto the expected output state. The quantum states were teleported over a distance of $6~\mathrm{mm}$ with a rate of $40\cdot10^3~\mathrm{events/s}$ which exceeds many reported implementations of teleportation.

Previously, we have implemented a teleportation protocol in superconducting circuits up to the single-shot measurement step \cite{Baur2012}. In that experiment three superconducting qubits were coupled to a single transmission line resonator which was used as a quantum bus~\cite{Majer2007} and for an averaged joint read-out of all qubits~\cite{Filipp2009b}. Full three-qubit quantum state tomography was performed at the last step to show that the resulting state of all three qubits is a genuine tripartite entangled state. In post-processing this state was projected onto the basis states of the two sender qubits to characterize the coherent part of the teleportation protocol \cite{Baur2012}.

\section{Experiments and Results}
%Experimental setup
In the realization of teleportation presented here, we use three superconducting transmon qubits~\cite{Koch2007} (Q1, Q2, Q3) coupled to three superconducting coplanar waveguide resonators (R1, R2, R3) in the circuit quantum electrodynamic setup~\cite{Wallraff2004} shown in Fig.~\ref{fig:schematicsample}. At the sender, qubits Q1 and Q2 are coupled capacitively to resonator R1, at the receiver Q3 is coupled to R3. In addition,  Q2 and Q3 are coupled to R2 to distribute entanglement between the two parties. We perform single qubit rotations by applying amplitude and phase controlled microwave pulses through individual charge gate lines. The transition frequency of each qubit is controlled by individual flux bias lines. The resonators act as quantum buses~\cite{Majer2007} to realize two-qubit controlled-{\sc phase} (\textsc{cphase}) gates~\cite{DiCarlo2009}.

\begin{figure*}[!t]
\includegraphics[width=0.98\textwidth]{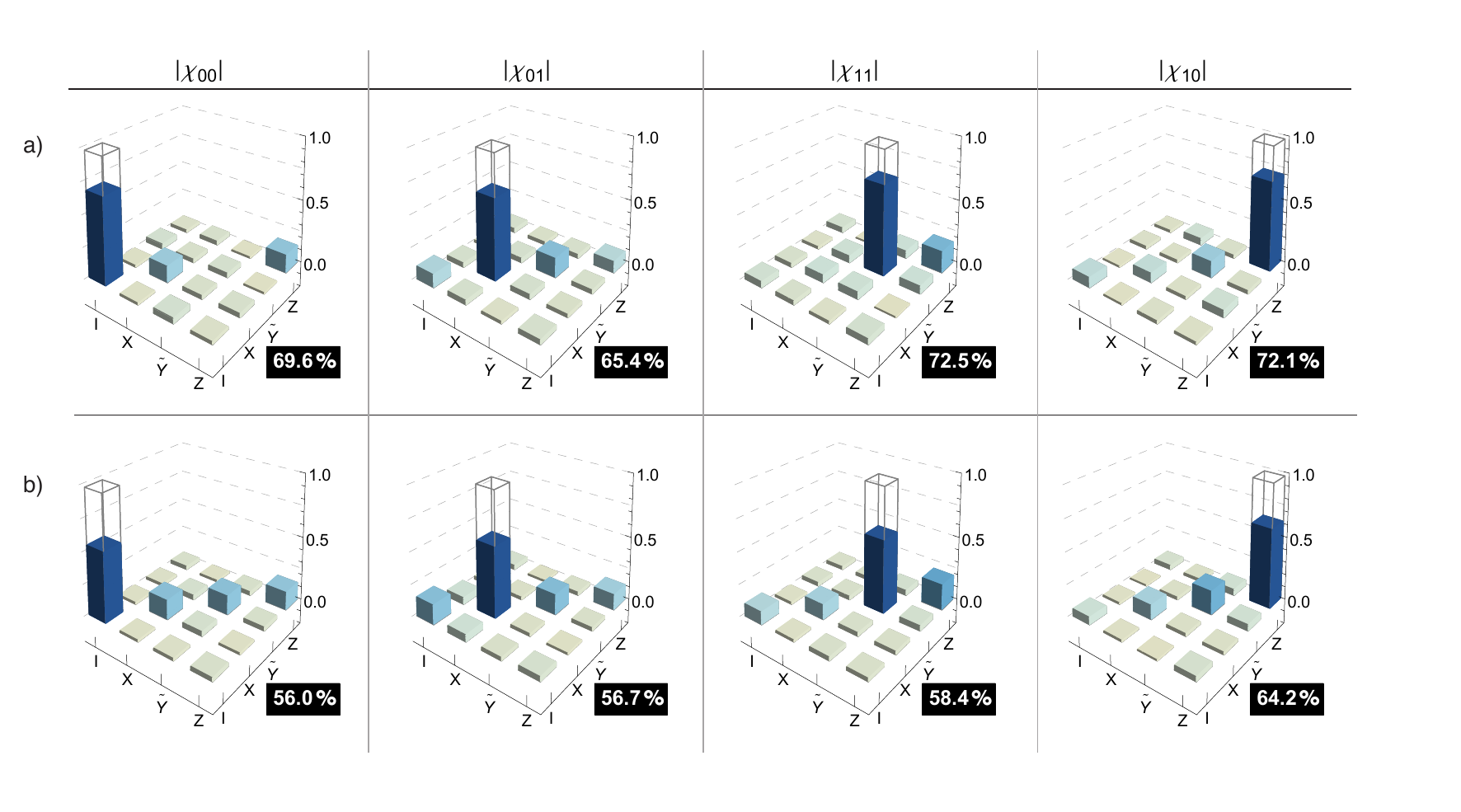}
\caption{Absolute values of the experimentally extracted process matrices $\left|\chi\right|$ describing the state transfer from Q1 to Q3 for the two different measurement schemes: a) Post-selected on one of the measurement outcomes $00,01,10,11$ using  phase sensitive detection, b) simultaneous detection of all four measurement outcomes using phase preserving detection. The process fidelities are indicated in the black boxes.}
\label{fig:chis}
\end{figure*}

In about $15\,\%$ of the experimental runs we find that one or more of the qubits are initially in a thermally excited state (see appendix). These experimental traces are disregarded in our analysis. Alternatively, the initialization efficiency could be improved by using active initialization schemes~\cite{Riste2012, Geerlings2012b}.

Using single qubit rotations and a {\sc cphase}-gate a Bell state with fidelity $93\,\%$ is created on demand between qubits Q2 and Q3 which is shared between the sender and the receiver, see blue box in Fig.~\ref{fig:pulses}b). The state of qubit Q1 to be teleported is then prepared using a single qubit rotation.

%Bell measurement
In our setup, projective qubit measurement is naturally performed in the computational basis $\ket{00},\ket{01}, \ket{10}, \ket{11}$~\cite{Blais2004,Filipp2009b}.
To effectively perform a Bell measurement one can map the Bell basis onto the computational basis using a \textsc{cnot}-gate and a Hadamard gate as suggested in Ref.~\cite{Brassard1998}, see red box in Fig.~\ref{fig:pulses}a). We realize this basis transformation in our system by using single qubit rotations and a \textsc{cphase}-gate. Then we perform a joint read-out \cite{Filipp2009b} of the states of Q1 and Q2 by measuring the transmission amplitude and phase of resonator R1.
A given Bell state $\left\{\ket{\Phi^-},\ket{\Psi^-},\ket{\Phi^+},\ket{\Psi^+}\right\}$ is transformed to the corresponding computational basis state $\left\{\ket{00},\ket{01},\ket{10},\ket{11}\right\}$ resulting in an output state $\left|\psi_\mathrm{out}\right\rangle = \left\{\mathds{1},\hat{\sigma}_x,\hat{\sigma}_z,i\hat{\sigma}_y\right\}\left|\psi_\mathrm{in}\right\rangle$ of Q3.
%\begin{align}
%\ket{\Phi^-}\Rightarrow&\ket{00}&\longrightarrow&&\ket{\psi_\mathrm{out}}&=\ket{\psi_\mathrm{in}}\nonumber\\
%\ket{\Psi^-}\Rightarrow&\ket{01}&\longrightarrow&&\ket{\psi_\mathrm{out}}&=\hat{\sigma}_x\ket{\psi_\mathrm{in}}\nonumber\\
%\ket{\Phi^+}\Rightarrow&\ket{10}&\longrightarrow&&\ket{\psi_\mathrm{out}}&=\hat{\sigma}_z\ket{\psi_\mathrm{in}}\nonumber\\
%\ket{\Psi^+}\Rightarrow&\ket{11}&\longrightarrow&&\ket{\psi_\mathrm{out}}&=i\hat{\sigma}_y\ket{\psi_\mathrm{in}}
%\label{eqn:bellmeas}
%\end{align}
Since the Bell state measurement has four randomly distributed measurement outcomes, high fidelity single-shot read-out is required to identify each of these outcomes.
In our setup this is accomplished by using a Josephson parametric amplifier~\cite{Castellanos2008, Vijay2011}.

%Methods
%Two different measurement schemes: 1. Post-selecting on only one state, 2. distinguish all four states.
A parametric amplifier can be operated in two different modes. In the phase sensitive mode~\cite{Castellanos2008} the amplifier has the highest gain and in principle adds no noise to one of the detected quadratures of the signal. In the phase preserving mode~\cite{Eichler2011a} the total gain is lower but both quadrature amplitudes of the detected electromagnetic field are amplified. In our parameter regime, these two modes allow for the possibility to perform a measurement and either post-select on only one of the four Bell states or to distinguish all four Bell states simultaneously with high fidelity.

%One state: Paramp 1 in phase sensitive mode, optimized contrast of $\left|00\right\rangle$ to the other states $\left|01\right\rangle,\left|10\right\rangle,\left|11\right\rangle$. Bell state is chosen by changing the phase of the $\pi/2$-pulse before the joint-measurement.
If the measurement of Q1 and Q2 returns $\ket{00}$, qubit Q3 is instantaneously projected to the desired state $\ket{\psi_\mathrm{in}}$ without the necessity for additional rotations.
This is achieved by operating the parametric amplifier connected to R1 in the phase sensitive mode~\cite{Castellanos2008} and optimizing the read-out contrast between the state $\ket{00}$ detected with a fidelity of $(90.8 \pm 0.3)\,\%$ and all other states $\ket{01},\ket{10},\ket{11}$ which are not distinguished with high fidelity from each other, see appendix.

With a second parametric amplifier a measurement tone transmitted through resonator R3 is used to readout the state of qubit Q3 with a single shot fidelity of $(87.9 \pm 0.9)\,\%$. State tomography of Q3 conditioned on a $\ket{00}$ measurement of Q1 and Q2 ideally occuring with a probability of $1/4$ reveals the original input state with an average fidelity of $\mathcal{\bar{F}}_s = (82.4 \pm 2.3)\,\%$, see Fig.~\ref{fig:states}. By characterizing $\ket{\psi_\mathrm{out}}$ for four linearly independent input states $\ket{\psi_\mathrm{in}}$, we perform full process tomography of the state transfer from Q1 to Q3 to reconstruct the process matrix $\chi_{00}$. The teleportation process is realized with a fidelity of $\mathcal{F}_p = (69.6\pm2.3)\,\%$ with respect to the expected identity operation.

We are able to map any of the Bell states to the computational basis state $\ket{00}$ on demand by applying $\pi$-pulses to Q1 and Q2 right before their joint read-out. This allows us to post-select \emph{individually} on any of the four Bell states and to determine the corresponding process matrices $\chi_{00,01,10,11}$.
The experimentally obtained process matrices are shown in Fig.~\ref{fig:chis}a) and agree well with the expected processes.
The average output state fidelity $\mathcal{\bar{F}}_s=(80.7\pm2.0)\,\%$ of all four processes is clearly above the classical limit of 2/3~\cite{Massar1995}. This results in an average process fidelity when post-selecting on a single Bell state of $\mathcal{F}_p=(69.9\pm2.0)\,\%$, well above the classical limit of $1/2$.
The process fidelity $\mathcal{F}_p$ is related to the average output state fidelity $\bar{\mathcal{F}}_s$ as $\mathcal{F}_p=({\mathcal{\bar{F}}_s(d+1)-1})/{d}$~\cite{Horodecki1999} where $d$ is the dimensionality of the input and output state.
The output state fidelity is predominantly limited by the relaxation and dephasing of our qubits which affects both the effective gate- and read-out fidelity (see appendix).

%All four states: Paramp 1 is operated in the phase preserving mode. Lower gain but both quadratures are amplified. Chose readout power and frequency such that all four states can be distinguished according to their position in the IQ-plane.
Operating the parametric amplifier in the phase preserving mode~\cite{Eichler2011a} and recording both quadrature amplitudes $(I, Q)$ of the measurement signal simultaneously we are also able to distinguish all four Bell states in one single measurement with a detection fidelity of $(80.7 \pm1.0)\,\%$.
Performing again state and process tomography we find an average output state fidelity of the transferred state of $\bar{\mathcal{F}}_s=(72.5 \pm 2.6)\,\%$ and an average process fidelity of $\mathcal{F}_p=(58.8\pm 2.4)\,\%$ above the classical limits of $2/3$ and $1/2$, respectively. The process matrices (Fig.~\ref{fig:chis}b) show prominently the characteristic features of the expected processes.
The lower fidelities obtained with this method compared to the previously described method are attributed to the lower detection fidelity.

\section{Prospects}
%Outlook
The recent increase in coherence times in single qubit circuits~\cite{Paik2011, Sandberg2012a} may lead to major improvements of the fidelity of single-shot read-out and single- and two-qubit gates also in more complex circuits such as the one presented here.
In addition longer coherence times facilitate the implementation of active feedback, which consists of a rotation of the output state conditioned on the outcome of the Bell measurement. This can be achieved using fast digital signal processing platforms~\cite{Riste2012b} such as field programmable gate arrays (FPGAs) which perform real-time data analysis to trigger the corresponding microwave pulses for the rotation of the qubit.
Incorporating these developments in next generation experiments will allow the demonstration of deterministic teleportation with feed-forward and is an important step towards the realization of quantum networks with superconducting circuits.

\acknowledgments{We acknowledge fruitful discussion with and feedback on the manuscript from Alexandre Blais, Florian Marquardt, and Stefan Filipp. We thank Yulin Liu for his contributions in early stages of the experimental work. This work was financially supported by ETH Zurich, the EU Integrated Project SOLID, and the SNF NCCR QSIT.}

%Supplementary material
\appendix

\section{Sample Parameters}
The sample consists of three superconducting coplanar waveguide resonators and three qubits of the transmon type~\cite{Koch2007} as depicted in Fig.~\ref{fig:schematicsample}. The resonators R1 and R3 have bare resonance frequencies
$\nu_\mathrm{r}=\{7.657, 9.677\}$~GHz, respectively. They are coupled by gap- and finger capacitors to their in- and output lines. The capacitances are designed such that the decay rate through the input port is approximately 100 times lower than through the output port providing the dominant decay channel for these resonators. The overcoupled resonator decay rates are measured to be
$\kappa/2\pi=\{2.4, 2.5\}$~MHz.
The resonator R2 is not coupled to any in- or output line. Its resonance frequency is approximately $8.7$~GHz and its decay rate is expected to be close to the internal decay rate 
%of $\kappa < 50$~kHz 
\cite{Goppl2008}.
From spectroscopic measurements we determine the maximum transition frequencies
$\nu_\mathrm{max}=\{6.273, 7.373, 8.390\}$~GHz and charging energies
$E_\mathrm{C}/h=\{0.297, 0.303, 0.287\}$~GHz of the qubits Q1, Q2, and Q3 respctively, where $h$ is Planck's constant.

Qubits Q1 and Q2 are coupled to resonator R1 with coupling strengths
$g/2\pi =\{0.260, 0.180\}$~GHz, and Q3 is coupled to resonator R3 with a coupling strength of
$g/2\pi =0.240$~GHz. The coupling of Q2 and Q3 to R2 is estimated from the transverse coupling strength (see below) to be $g/2\pi = 0.2$~GHz each.

For the presented experiments, the qubits were tuned to transition frequencies
$\nu=\{5.114, 6.004, 6.766\}$~GHz with miniature superconductiong coils mounted underneath the chip~\cite{Fink2009}. At these frequencies we have determined their energy relaxation times to be $T_1=\{2.6, 2.4, 2.0\}~\mu\mathrm{s}$ and coherence times  $T_2=\{1.8, 1.4, 1.2\}~\mu\mathrm{s}$.

%Pulse scheme
\section{Pulse Scheme}
All biased qubits and resonators are separated in frequency from each other by at least $800$~MHz to suppress cross-talk.

\begin{figure}
\includegraphics{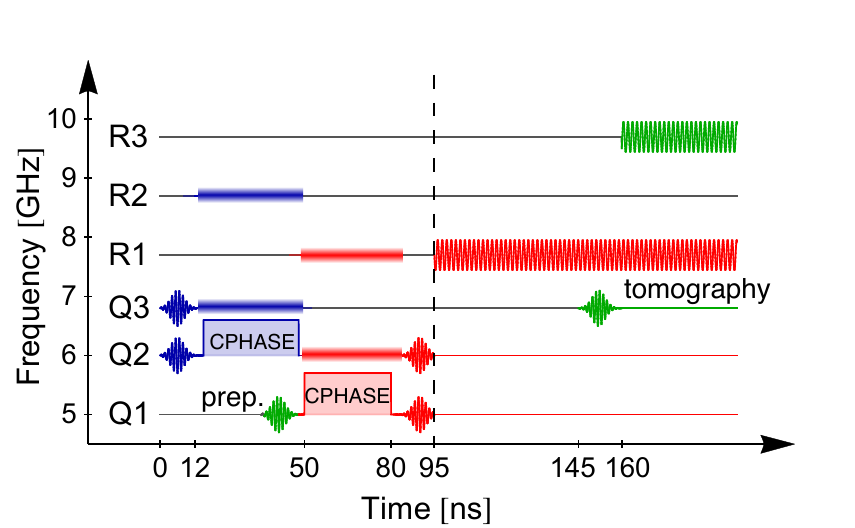}
\caption{Pulse sequence of the teleportation protocol. The pulses implement the creation of an entangled pair between Q2 and Q3 (blue), the preparation of the state to be teleported on Q1 (green), the basis transformation from the Bell to the computational basis and the subsequent readout of Q1 and Q2 (red), and the state tomography of Q3 (green). Gaussian shaped sinusoids represent the microwave pulses applied to the different charge bias lines of the qubits, sinusoids on the resonators represent the readout tones, and the squares labelled ``\textsc{cphase}" represent the flux pulses that shift the frequency of a qubit to implement a controlled-\textsc{phase} gate between the marked qubits, where the interaction is mediated through the resonator indicated with a bar of the same color as the flux pulse.}
\label{fig:pulsescheme}
\end{figure}

The protocol shown in Fig.~\ref{fig:pulses}b) is implemented with the pulse scheme depicted in Fig.~\ref{fig:pulsescheme}. Single qubit rotations are implemented by $12$~ns long resonant gaussian-shaped DRAG~\cite{Motzoi2009, Gambetta2011a} microwave pulses. The controlled-\textsc{phase} gate is implemented by shifting the qubits with fast magnetic flux pulses to the avoided level crossing between the $\ket{11}$ and $\ket{02}$ states of the involved qubits \cite{Strauch2003, DiCarlo2009}. The transverse coupling strengths of $J^{\mathrm{Q1,Q2}}_{11,02}/2\pi=17$~MHz (between qubits Q1 and Q2) and  $J^{\mathrm{Q2,Q3}}_{11,02}/2\pi=13$~MHz (between Q2 and Q3) lead to pulse lengths for the \textsc{cphase} gates of $t=\{29.5, 37.3\}$~ns, respectively.

The Bell measurement (Fig.~\ref{fig:pulsescheme}, red elements) allows to map any of the four Bell states to the $\ket{00}$ state by adding $\pi$-pulses to Q1 and Q2 right before the measurement tone. The $\pi$-pulses are implemented by changing the phases of the preceding $\pi/2$-pulses which is equivalent to adding separate pulses.

%Histogram
\section{Qubit Read-Out}
In order to realize single-shot measurement, the output signals of resonators R1 and R3 are amplified by individual Josephson parametric amplifiers \cite{Castellanos2008, Vijay2011}. The parametric amplifiers are similar to the one used in Ref.~\cite{Eichler2011a}. They are realized as $\lambda/4$ coplanar waveguide resonators terminated by an array of eleven superconducting quantum interference devices (SQUIDs) which provide the necessary non-linearity and make the operation frequencies tunable by miniature superconducting coils on the bottom of the sample holder. The maximum frequencies for the two parametric amplifiers are $\nu_\mathrm{max}=\{8.349, 10.141\}$~GHz for R1 and R3 respectively. In order to provide a fast response to the provided input signals, high input capacitors for the resonators were fabricated which result in a measured linewidth of $\kappa/2\pi =\{334, 548\}$~MHz in the linear regime.
For the experiments they were tuned to have the maximum gain of $G=\{23.9, 23.5\}$~dB  with a $3$~dB bandwidth of $B/2\pi=\{12, 55\}$~MHz at the frequencies $\nu_\mathrm{exp}=\{7.693, 9.712\}$~GHz.

\begin{figure}
\includegraphics[width=0.45\textwidth]{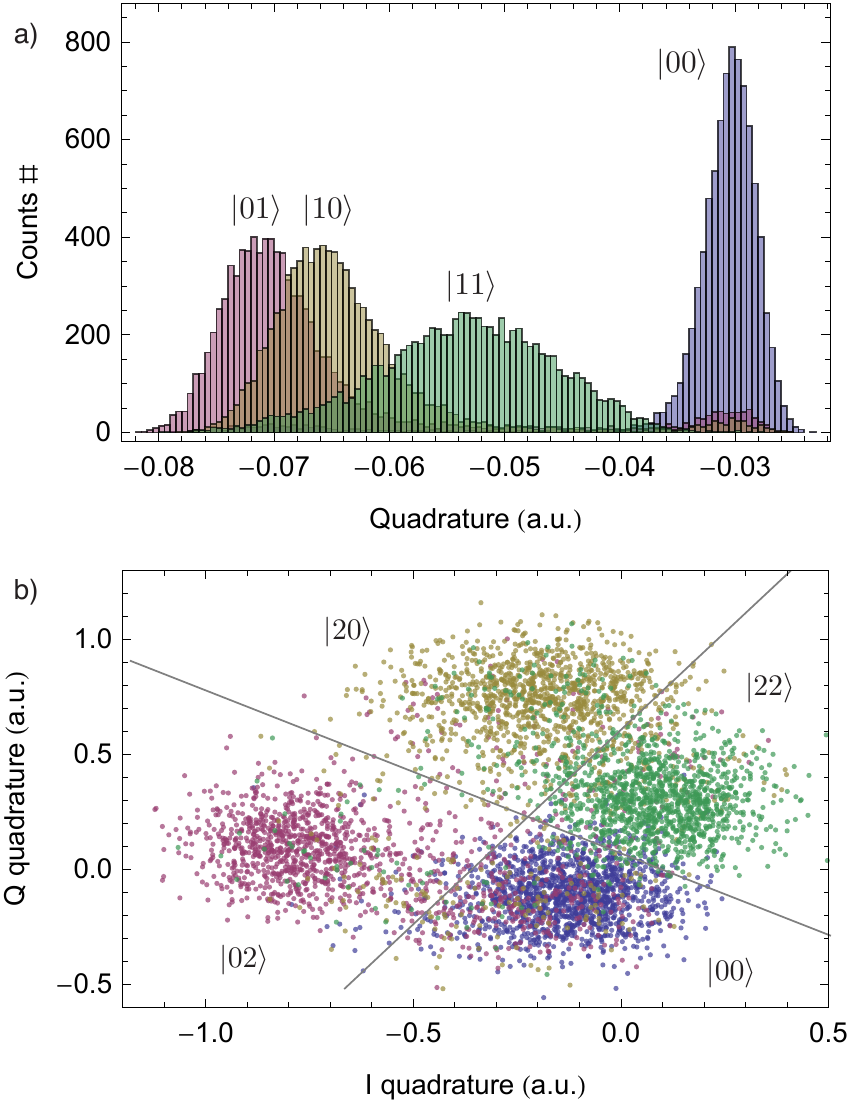}
\caption{a) Histogram of the integrated signal quadrature amplitude amplified phase sensitively when preparing the states $\ket{00}$ (blue), $\ket{01}$ (red), $\ket{10}$ (yellow), and $\ket{11}$ (green).
b) Histogram of integrated $(I, Q)$ quadratures of the measurement signal amplified in the phase preserving mode when preparing the states $\ket{00}$ (blue), $\ket{02}$ (red), $\ket{20}$ (yellow), and $\ket{22}$ (green). The chosen thresholds  are indicated with gray lines.}
\label{fig:histograms}
\end{figure}

For the experiments in which we post-select on an individual Bell-state, the transmission of R1 was measured at the readout frequency $\nu_\mathrm{ro}=7.693$~GHz which is the mean value of the effective resonator frequencies for the qubits Q1 and Q2 in the state $\ket{00}$ and $\ket{01}$. The parametric amplifier is used in the phase-sensitive mode by tuning its transition frequency such that the maximum gain was achieved at the readout frequency $\nu_\mathrm{ro}$ where it was also pumped.
Preparing the four computational basis states  $\ket{00},\ket{01}, \ket{10}, \ket{11}$, applying a measurement tone at R1 and integrating the amplified transmission signal for $250$~ns results in a distribution of the acquired signals as shown in Fig.~\ref{fig:histograms}a). Since we optimized for the readout contrast between the  $\ket{00}$ and $\ket{01}$ states, the mean values of the distributions of the integrated signals for these two states (blue and red bars in Fig.~\ref{fig:histograms}a) have the largest separation. However due to the finite qubit lifetime, some of the $\ket{01}$ and $\ket{10}$ states decay into the ground state and are visible in the analysis as such. We choose a threshold for the integrated quadrature values to discriminate  $\ket{00}$ from all other basis states $\ket{01}, \ket{10}, \ket{11}$ with a fidelity of $(90.8 \pm 0.3)\,\%$.

\begin{figure*}[!t]
\includegraphics[width=0.59\textwidth]{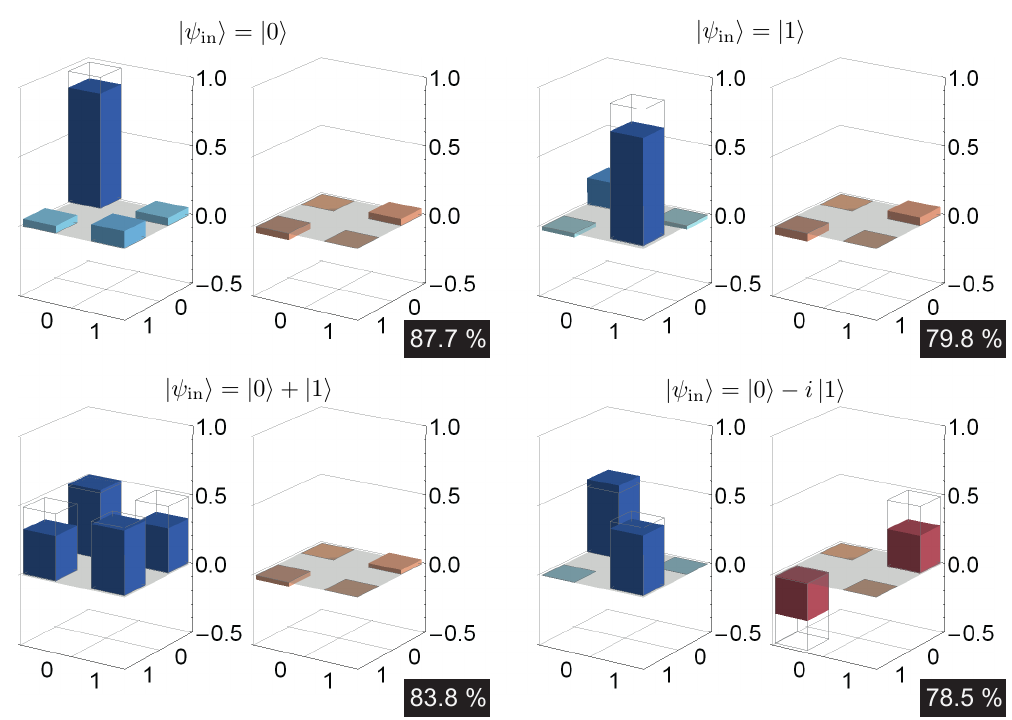}
\caption{Real (blue) and imaginary (red) parts of the reconstructed density matrices of the state $\ket{\psi_\mathrm{out}}$ for the indicated input states $\ket{\psi_\mathrm{in}}$ obtained from state tomography when post-selecting data on a $\ket{00}$ outcome of the Bell measurement. The ideally expected outcomes are indicated with wireframes. The state fidelities are indicated in the black boxes.}
\label{fig:states}
\end{figure*}

In the experiments in which we are able to distinguish between all four Bell states simultaneously, the readout frequency $\nu_\mathrm{ro}=7.685$~GHz is chosen to be the mean of the effective resonator frequencies for the qubits Q1 and Q2 in the state $\ket{01}$ and $\ket{10}$. The parametric amplifier is used in the phase-preserving mode by detuning the pump frequency $6.25$~MHz from the readout frequency. In this way the gain $G=20$~dB at the readout frequency and the effective bandwidth are smaller than in the previous case, but both quadratures of the electromagnetic field are amplified. By preparing the computational basis states and recording the integrated transmitted signals of both quadratures $(I, Q)$ simultaneously, we can map every measurement outcome to a point on the complex plane. By adjusting the pump power and frequency as well as the readout power we find settings which maximize the distinguishability of all four states by their location on the complex plane. Applying a
resonant pulse between the $\ket{1}$ and $\ket{2}$ state just before the readout increases the contrast between the different states further. By choosing thresholds that divide the complex plane into four sectors (see Fig.~\ref{fig:histograms}b) and assigning the corresponding state to all measurement outcomes in a certain area, we identify $(80.7 \pm 1.0)\,\%$ of the prepared states correctly.

\section{Efficiency}
In every experiment we apply a $500$~ns long measurement tone before the beginning of the protocol to both resonators in order to verify that all the qubits are initialized in their ground states. When using the parametric amplifier in the phase-sensitive mode, we find that in about $15\%$ of the cases, one or more qubits are in the excited state before the protocol. These measurements are then not considered for the data analysis, since the protocol did not start from the desired initial state. When the parametric amplifier is used in the phase-preserving mode the distinction between the ground and excited states in this measurement is achieved by defining a threshold forming a circle in the complex plane separating ground state measurements from thermally excited state measurements. Optimizing for a high output state fidelity we chose a threshold which discards $30\%$ of all measurement runs, i.e.~those starting in an excited initial state.

We emphasize once more that $85\%$ ($70\%$) of all runs of the teleportation protocol, i.e. those with qubits initialized in the ground state, are used in the analysis of the phase sensitive (preserving) measurements. This is in contrast to experiments in which optical photons are used as qubits, where the maximal reported efficiency is $0.1\%$~\cite{Nolleke2012}. Using active initialization schemes which have been demonstrated for superconducting circuits~\cite{Riste2012, Geerlings2012b}, the efficiency can likely be improved to approach $100\%$ for sufficiently long qubit coherence times.

\section{State- and Process Tomography}
In order to characterize the process describing the state transfer from Q1 to Q3 we performed full process tomography. By performing state tomography on the output state $\ket{\psi_\mathrm{out}}$ for four different input states $\ket{\psi_\mathrm{in}}=\ket{0},\ket{1}, (\ket{0}+\ket{1})/\sqrt{2}, (\ket{0}-i\ket{1})/\sqrt{2} $ (see Fig.~\ref{fig:states}), we obtain the process matrix $\chi$ through linear inversion.
For state tomography we measure a repeatedly prepared state with different measurement operators. This is implemented by applying either no pulse, a $\pi/2$-pulse about the $x$- or $y$-axis or a $\pi$-pulse to the qubit just before measurement.

\section{Error Budget}
The finite coherence and dephasing times of our qubits are a source of error which limit the output state fidelity. The fidelity of the measurement of Q3 through R3 affects directly the state fidelity of $\ket{\psi_\mathrm{out}}$. From the measured probabilities of correctly identifying the states $\ket{0}$ and $\ket{1}$ on Q3 we calculate the limit of the output state fidelity through this source of error to be $\mathcal{\bar{F}}_s= 94\,\%$. In addition, the misidentification of the Bell states leads to an effective dephasing of $\ket{\psi_\mathrm{out}}$. This limits the fidelity further to $\mathcal{\bar{F}}_s= 89\,\%$ and $\mathcal{\bar{F}}_s= 82\,\%$ for the case in which we post-select on one Bell state only and in which we distinguish all Bell states with each measurement respectively.
Since both of these numbers are about $10\,\%$ higher than the actually measured fidelities, it is plausible to assign the remaining errors to the limited gate fidelities.
Determining the gate errors independently shows that we perform single qubit operations with a fidelity greater than $98\,\%$ and the creation of a Bell state with a fidelity of $93\,\%$.

\bibliographystyle{apsrev4-1}
\bibliography{teleportation}

\end{document}